\documentclass[twocolumn, letter, superscriptaddress]{revtex4} 
  
\usepackage{amsmath, array} 
\usepackage{graphicx}
\usepackage{hyperref} 
\usepackage{color}
\usepackage[normalem]{ulem}
\usepackage{soul,xcolor}

\setstcolor{red}

\newcommand{\EqLabel}[1]{\label{#1}} 

\newcommand{\revision}[1]{{\color{black} #1}}

\newcommand{\john}[1]{{\color{black} #1}}

\begin{document}

\title{
Phonon-mediated repulsion, sharp transitions and (quasi)self-trapping in the extended Peierls-Hubbard model 
}

\author{J. Sous} \thanks{Correspondence to jsous@physics.ubc.ca} \affiliation{\!Department \!of \!Physics and
  Astronomy, \!University of\!  British Columbia, \!Vancouver, British
  \!Columbia,\! Canada,\! V6T \!1Z1} \affiliation{\!Stewart Blusson Quantum Matter \!Institute, \!University
  of British Columbia, \!Vancouver, British \!Columbia, \!Canada,
  \!V6T \!1Z4} 
\author{M. Chakraborty} \affiliation{Department of Physics, Indian Institute of Technology, Kharagpur, India}
\author{C. P. J. Adolphs} \affiliation{\!Department \!of \!Physics and
  Astronomy, \!University of\!  British Columbia, \!Vancouver, British
  \!Columbia,\! Canada,\! V6T \!1Z1} \affiliation{\!Stewart Blusson Quantum Matter \!Institute, \!University
  of British Columbia, \!Vancouver, British \!Columbia, \!Canada,
  \!V6T \!1Z4} 
\author{R. V. Krems}
\affiliation{\!Department \!of Chemistry, \!University
  of\!  British Columbia, \!Vancouver, British \!Columbia,\! Canada,\!
  V6T \!1Z1} 
\author{M. Berciu}
\affiliation{\!Department \!of \!Physics and Astronomy, \!University
  of\!  British Columbia, \!Vancouver, British \!Columbia,\! Canada,\!
  V6T \!1Z1} \affiliation{\!Stewart Blusson Quantum Matter \!Institute, \!University
  of British Columbia, \!Vancouver, British \!Columbia, \!Canada,
  \!V6T \!1Z4}
  
\begin{abstract}
We study two identical fermions, or two hard-core bosons, in an
infinite chain and coupled to phonons by interactions that modulate
their hopping as described by the Peierls/Su-Schrieffer-Heeger (SSH)
model. We show that exchange of phonons generates effective
nearest-neighbor {\it repulsion} between particles and also gives rise
to interactions that move the pair as a whole. The two-polaron phase
diagram  exhibits two sharp transitions, leading to
light dimers at strong coupling and the flattening of the dimer
dispersion at some critical values of the parameters. This dimer
(quasi)self-trapping occurs at coupling strengths where single
polarons are mobile. This illustrates that, depending on the strength
of the phonon-mediated interactions, the coupling to phonons may
completely suppress or strongly enhance quantum transport of
correlated particles.
\end{abstract}
\date{\today}

\maketitle

{\em Introduction}: Strongly correlated quantum materials exhibit rich
physics with many features yet to be understood. Correlated lattice
systems are modeled by the extended Hubbard model, which includes
inter-site interactions giving rise to interesting physics such as
superfluid - Mott insulator transitions \cite{mot}, antiferromagnetism
\cite{af, af-2}, high-Tc superconductivity \cite{highTc}, twisted
superfluidity \cite{twisted}, supersolids \cite{ss}. However, the
extended Hubbard model does not include interactions with phonons,
which are essential for quantum materials. Here, we show that the
interplay of the extended Hubbard interactions with phonon-mediated
couplings leads to new unique features, such as self-trapping of
correlated pairs and the formation of light (mobile) dimers
in the regime of strong interactions, both between the particles and
with phonons.

A particle (electron, exciton, etc.) dressed with phonons is a
polaron. If phonons modulate the on-site energy of the particle, as is
the case for electrons in ionic lattices, polarons can be viewed as the
bare particle dragging a cloud of phonons. Such polarons are
always heavier than the bare particle
\cite{Landau,Feynman1,Feynman2,Migdal,Eliash,Holstein1,Holstein2,Froh1,Froh2}. On the other hand, phonons
also modulate the hopping of the particle between sites. Such interactions are
important for electrons in conjugated polyenes, where they are
described by the Su-Schrieffer-Heeger (SSH) model \cite{SSH1a,SSH1b,SSH1c,SSH2a,SSH2b}, or
for excitons in molecular solids, where they are described by the
Peierls model. Polarons arising from the SSH/Peierls interactions
exhibit sharp transitions \cite{Dominic} into strong-coupling regimes
where the polaron (dressed particle) is {\it lighter} than the bare particle
\cite{Dominic,Edwards1,Edwards2,Edwards3,Edwards4,Trugman-tJ}.

The interplay of the SSH/Peierls couplings and the extended Hubbard
interactions may alter the behaviour of strongly correlated quantum
systems. For example, in the limit of half-filling, an interplay of
phonon-mediated attraction with repulsive Hubbard interactions is
known to lead to a competition between the Mott-insulator and
Peierls-insulator phases \cite{pearson}. Here, we consider polarons
arising in the two-particle limit of an extended Hubbard model coupled
to phonons through the SSH/Peierls couplings. This is critical for
understanding quantum transport of interacting excitons in devices
based on organic semiconductors (such as low-temperature solar cells)
\cite{materials, solar} and the prospects of observing the
Mott-insulator/Peierls-insulator competition with highly controllable
ultracold atoms/molecules systems, which require understanding of emergent interactions in the few-particle limit.
The extended Peierls-Hubbard model
can be realized for hard-core bosons with ions in rf-traps
\cite{ions1,ions2,ions3,ions4}, Rydberg atoms exchanging excitations \cite{rydbergs1,rydbergs2,rydbergs3,rydbergs4,rydbergs5},
self-assembled ultracold dipolar crystals \cite{dipolar-crystals1,dipolar-crystals2,dipolar-crystals3},
arrays of polar molecules trapped in optical lattices
\cite{polar-molecules1,polar-molecules2}, arrays of superconducting qubits
\cite{d-wave1,d-wave2,d-wave3,d-wave4,d-wave5,d-wave6}, and J-aggregates \cite{jagg}. Similar physics may also
arise in the context of interacting impurities in a Fermi degenerate
gas \cite{fdg1,fdg2,fdg3} or Bose-Einstein condensates \cite{bec1,bec2,bec3,bec4,bec5,bec6} of ultracold
atoms. Motivated by these experiments, we consider identical
fermions/hard-core bosons and show that the interplay between particle
statistics, particle interactions and coupling to phonons leads to
unique features such as {\it phonon-mediated repulsion} and {\it sharp
  transitions} in the ground-state properties of dimers including one
suggestive of {\it self-trapping}.

{\em Model:} We consider two identical fermions (fermionic atoms in the same
internal state), or equivalently, two hard-core bosons, placed in an
infinite chain \cite {statistics-1,statistics-2} described by the Hamiltonian ${\cal H}={\cal H}_{\rm
  p} + {\cal H}_{\rm ph} + \hat{V}$, where:
\begin{eqnarray}
{\cal H}_{\rm p} =-t \sum_{i}^{}\left( c_i^\dagger c_{i+1} +
h.c.\right) + U \sum_{i}^{}\hat{n}_i \hat{n}_{i+1}
\end{eqnarray}
is the extended Hubbard model of the bare particles with infinite
on-site repulsion, ${\cal H}_{\rm ph} = \Omega \sum_{i}^{}
b_{i}^\dagger b_i$ is the phonon Hamiltonian  (in units of $\hbar = 1$), and
\begin{equation}
\EqLabel{4} \hat{V}=g\sum_{i}^{}\left( c_i^\dagger c_{i+1} +
h.c.\right)\left( b_i^\dagger+b_i - b_{i+1}^\dagger-b_{i+1}\right)
\end{equation}
is the Peierls/SSH particle - phonon coupling \cite{Dominic}. Here,
$i$ is the site index, $\hat n_i = c_i^\dagger c_i$, $U$ is the
strength of the bare  nearest-neighbor (NN)
interactions and $\Omega$ is the phonon frequency. We characterize the
particle - phonon  effective coupling by the
dimensionless parameter $\lambda = 2g^2/(\Omega t)$.

{\em Methods:} We use two methods to investigate this problem. The
first is variational exact diagonalization (VED), a well-established,
unbiased numerical method, where the variational basis set is expanded
systematically, starting from the Bloch state for two adjacent
particles and zero phonons \cite{BKT,BT,M1}. The second method is
based on the Momentum Average (MA) approximation, a quasi-analytical
variational method that has been shown to be accurate for polarons
\cite{MA1,MA2,MA3}, including SSH polarons \cite{Dominic}. Here, we generalize
MA to study bound dimers  by allowing a variational space
where the two particles are either in adjacent sites or two sites
apart, interacting with a phonon cloud spread over at most three
adjacent sites (for more details, see the Supplementary Information);
we comment more on these choices below.

{\em Results}: First, we set $U=0$ and study whether exchange of
phonons suffices to bind two SSH polarons into a bipolaron. For
reference, we note that equivalent 1D models with long(er)-range
on-site energy-modulating couplings, such as the screened and unscreened Fr\"ohlich
couplings, show the appearance of stable bipolarons; for on-site
Holstein coupling, such bipolarons do not form \cite{BT,HK,M1}.

We find that for $U=0$, bipolarons do not form for any
coupling $\lambda$. To understand the implications of this result,
note that the bare particles ($\lambda =0$) bind only for $U \le -2t$.
This attraction is needed to compensate for the loss of kinetic energy
\cite{MBmp}. The SSH polaron dispersion --
and hence its kinetic energy -- remains significant at all particle -
phonon couplings. Our results thus show that the phonon-mediated
interaction is insufficient to compensate for the kinetic energy that
would be lost upon binding.

\begin{figure}[t]
  \centering \includegraphics[width=\columnwidth]{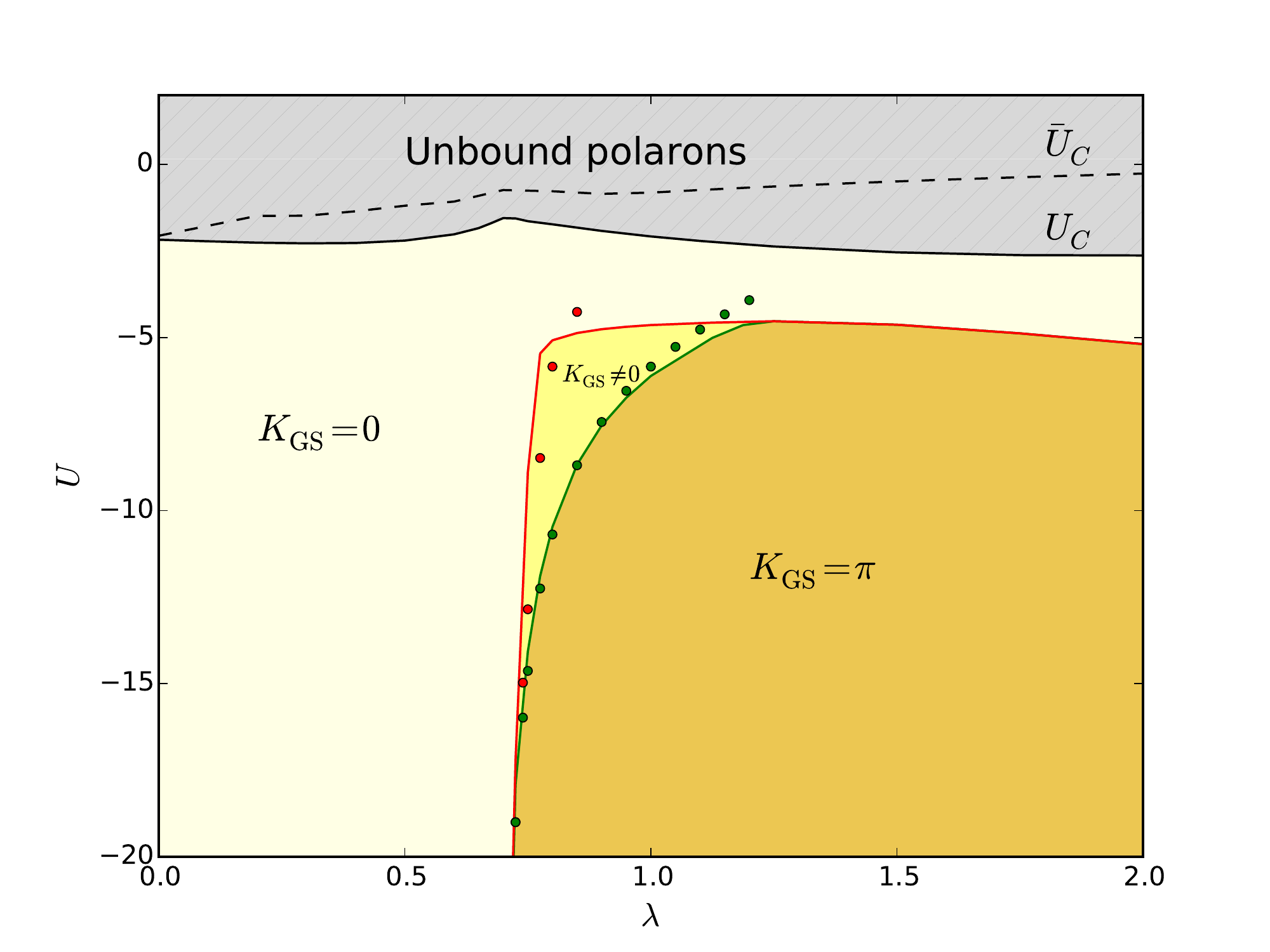} \caption{(color
    online) Two-polaron phase diagram at $t=1, \Omega=3$. The solid
    black line shows $U_C(\lambda)$ below which stable bound states
    form, while the {dashed} line shows $\bar{U}_C(\lambda)$; the
    difference between the two is the strength of the phonon-mediated
    {interaction}. Note that $U_C < \bar{U}_C$, which means that this
    interaction is repulsive. The red and green lines mark the sharp
    transitions of the bound dimer's GS. The lines are the VED results and
    the corresponding symbols are the MA results.  \label{fig1}} \end{figure}

To characterize quantitatively this phonon-mediated interaction, we
compute the values of $U = U_{C}(\lambda)$ corresponding to the onset
of stable bound state (we define the bound dimer to be stable if its ground
state energy is below the two-polaron continuum). 
We then compare $U_C(\lambda)$ with
$\bar{U}_C(\lambda)$, defined as the NN attraction needed to bind two
hard-core particles with dispersions identical to those of single SSH
polarons. This latter model mimics the renormalization of the
dispersion due to each particle creating and interacting with its own
cloud of phonons, but excludes the effective interactions due to
phonon exchange between the clouds. The phonon exchange occurs in the
full model, so $|\bar{U}_{C}(\lambda)| - |{U}_C(\lambda)|$ is an
estimate of the phonon-mediated NN attraction between polarons. Figure \ref{fig1} shows that $|U_{C}(\lambda)| > |\bar{U}_C(\lambda)|$ for all
$\lambda$. This means that the
phonon-mediated interaction is in fact {\it strongly repulsive}, in
stark contrast to what is observed for conventional polaron models \cite{BT,HK,M1}.

\john{
This surprising result can be  explained by considering the
limit $\Omega \gg |t|, |g|$  within perturbation theory (details in Supplementary Information).  Projecting out
high-energy states with one or more phonons, the effective
Hamiltonian for a single polaron becomes 
\begin{eqnarray}
\hat{h}_1 = \sum_{i}^{} \left( -t
c_i^\dagger c_{i+1}+ t_2 c_i^\dagger c_{i+2} + h.c.\right) -\epsilon_0
\sum_{i}^{}\hat{n}_i.
\end{eqnarray}
 The small cloud that forms in this limit does not
renormalize the NN hopping but it mediates a next-nearest-neighbor (NNN)
hopping $t_2=g^2/\Omega=\lambda t/2$ through the process $c^\dagger_i|0\rangle
\stackrel{\hat{V}}{\Longrightarrow}c^\dagger_{i+1} b^\dagger_{i+1}|0\rangle
\stackrel{\hat{V}}{\Longrightarrow} c^\dagger_{i+2} |0\rangle$. The four
processes $c^\dagger_i|0\rangle
\stackrel{\hat{V}}{\Longrightarrow}c^\dagger_{i\pm 1} b^\dagger_{i\pm
1}|0\rangle \stackrel{\hat{V}}{\Longrightarrow} c^\dagger_{i} |0\rangle$ and
$c^\dagger_i |0\rangle\stackrel{\hat{V}}{\Longrightarrow}c^\dagger_{i\pm 1}
b^\dagger_{i}|0\rangle \stackrel{\hat{V}}{\Longrightarrow} c^\dagger_{i}
|0\rangle$ explain the polaron formation energy $\epsilon_0= 4g^2/\Omega$
\cite{Dominic}.  The resulting polaron dispersion $E_P(k)= - \epsilon_0 - 2t
\cos (k) + 2 t_2 \cos (2k)$ is dominated by the  NNN hoping at large $\lambda$;
this explains both the transition, at $\lambda = 1/2$, of the polaron ground
state (GS) momentum from $k=0$ to a finite value that smoothly goes to
$k=\pi/2$, and why the polaron remains light at large $\lambda$ (for more
discussion, see Ref.  \cite{Dominic}).

Repeating the calculation for two particles, we find the corresponding
effective Hamiltonian to be
\begin{eqnarray}
\hat{h}_2 =\hat{h}_1 + \epsilon_0 \sum_{i}^{}
\hat{n}_i\hat{n}_{i+1}, 
\label{repulsion}
\end{eqnarray}
 illustrating the appearance of  phonon-mediated NN
{repulsion}. Its origin can be explained as follows: if the polarons are
$\delta \ge 2$  sites apart, each lowers its energy by $\epsilon_0$ through
hops to its adjacent sites and back, accompanied by virtual phonon emission
and absorption, as explained above. However, if the polarons are on adjacent
sites, then Fermi statistics blocks half of these processes, {\it i.e.} each
particle can only lower its energy by $\epsilon_0/2$. The energy cost for
polarons to be adjacent is, thus, $\epsilon_0=2\lambda t$.
}

It is very important to note that $\hat{h}_2$ also includes 
terms such as $c^\dagger_ic^\dagger_{i+1} |0\rangle
\stackrel{\hat{h}_2}{\Longrightarrow}
c^\dagger_{i+1}c^\dagger_{i+2}|0\rangle$. However, NNN hopping of one particle past the other is forbidden by
statistics (the particle at $i$ cannot emit a phonon and move to
$i+1$ because that site is occupied). Instead, these terms describe both
particles moving through $c^\dagger_ic^\dagger_{i+1} |0\rangle
\stackrel{\hat{V}}{\Longrightarrow}c^\dagger_ib^\dagger_{i+1}
c^\dagger_{i+2} |0\rangle \stackrel{\hat{V}}{\Longrightarrow}
c^\dagger_{i+1}c^\dagger_{i+2}|0\rangle$.  In other words, instead of
one particle hopping over the other, which is forbidden, each particle
moves by one site and a phonon is exchanged in the process. Thus, this
term is also a phonon-mediated effective interaction which would be
absent if phonons could not be exchanged between particles. In the
large $\Omega$ limit it happens to precisely compensate for the NNN
hopping forbidden by the particles' statistics, but that is not likely
to be the case throughout the parameter space. This shows that the
functional form of the effective phonon-mediated interaction must also
contain such ``pair-hopping'' terms in addition to the NN repulsion.
Such terms do not appear in models where phonons modulate the on-site particle energy (e.g., Holstein and Fr\"{o}hlich models).

\begin{figure}[t]
  \centering \includegraphics[width=\columnwidth]{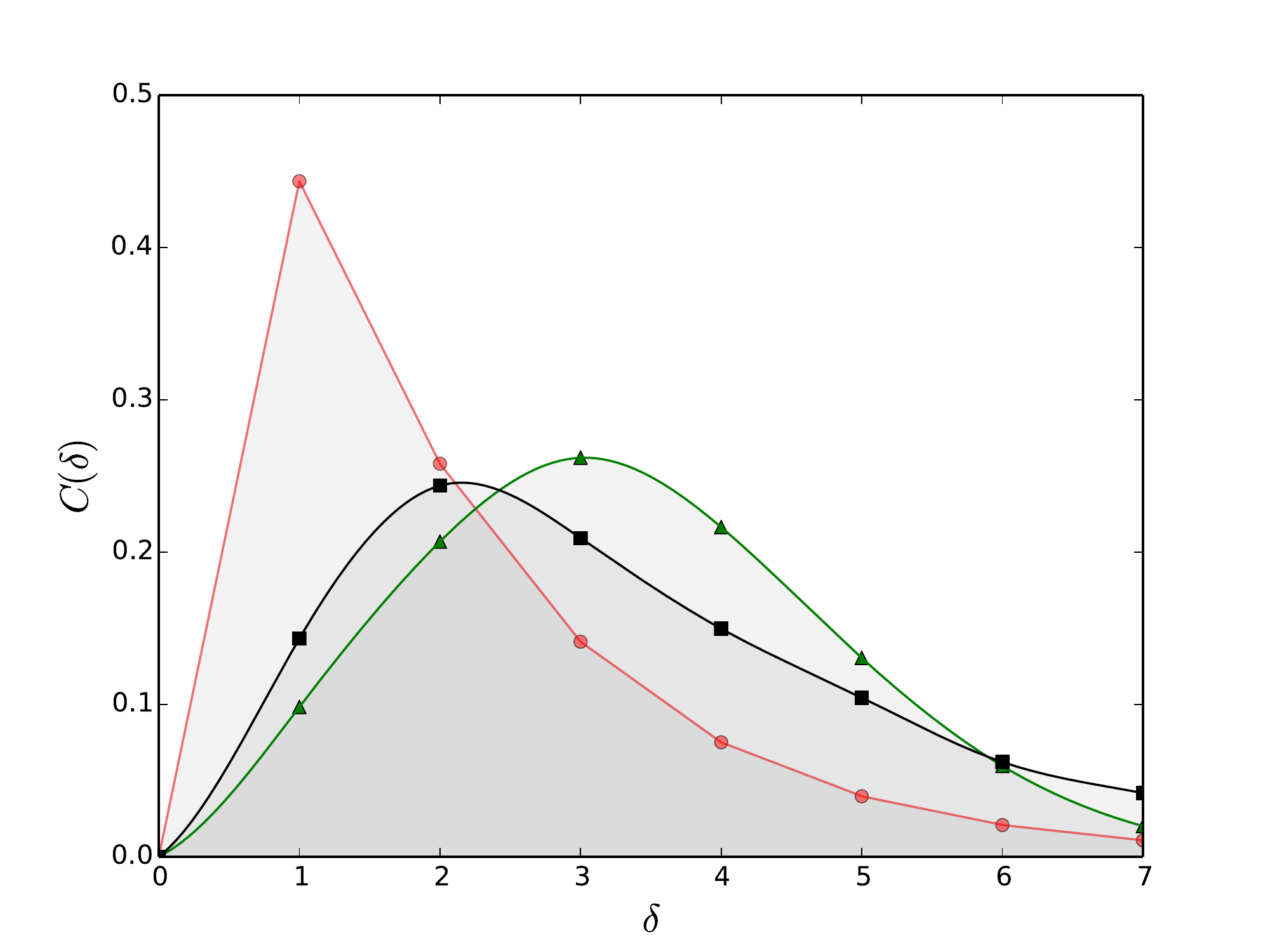} \caption{(color online) Correlation between the two particles $C(\delta) = \langle \Psi_{GS}| {1\over N} \sum_{i}^{}
\hat{n}_i\hat{n}_{i+\delta}|\Psi_{GS}\rangle$ {\em vs} separation $\delta$  for  $U=  U_C(\lambda) - 0.5$, $t=1, \Omega = 3$ and  $\lambda=0.1$ (red circles), $\lambda=0.7$ (green triangles), $\lambda=2.0$ (black squares).
\label{fig2}} \end{figure}

For smaller values of $\Omega$, the phonon clouds have more phonons
and are more extended spatially, and thus can mediate longer-range
effective interactions and hopping. Indeed, as shown in Fig.
\ref{fig2} for $\Omega = 3$ and $U = U_C(\lambda) - 0.5$, {\it i.e.}
just inside the dimer stability region, the bound particles favor
adjacent locations only for $\lambda \rightarrow 0$. At moderate and
strong couplings they are found with highest probability to be 2 or
even 3 sites apart, even though the bare attraction is NN only. This
suggests that the strong phonon-mediated NN repulsion is supplemented
by longer range effective attraction, and/or that binding is due to
kinetic energy gained through phonon-mediated ``pair-hopping'' terms
such as the one discussed above.

\begin{figure}[t]
    \centering \includegraphics[width=0.9\columnwidth]{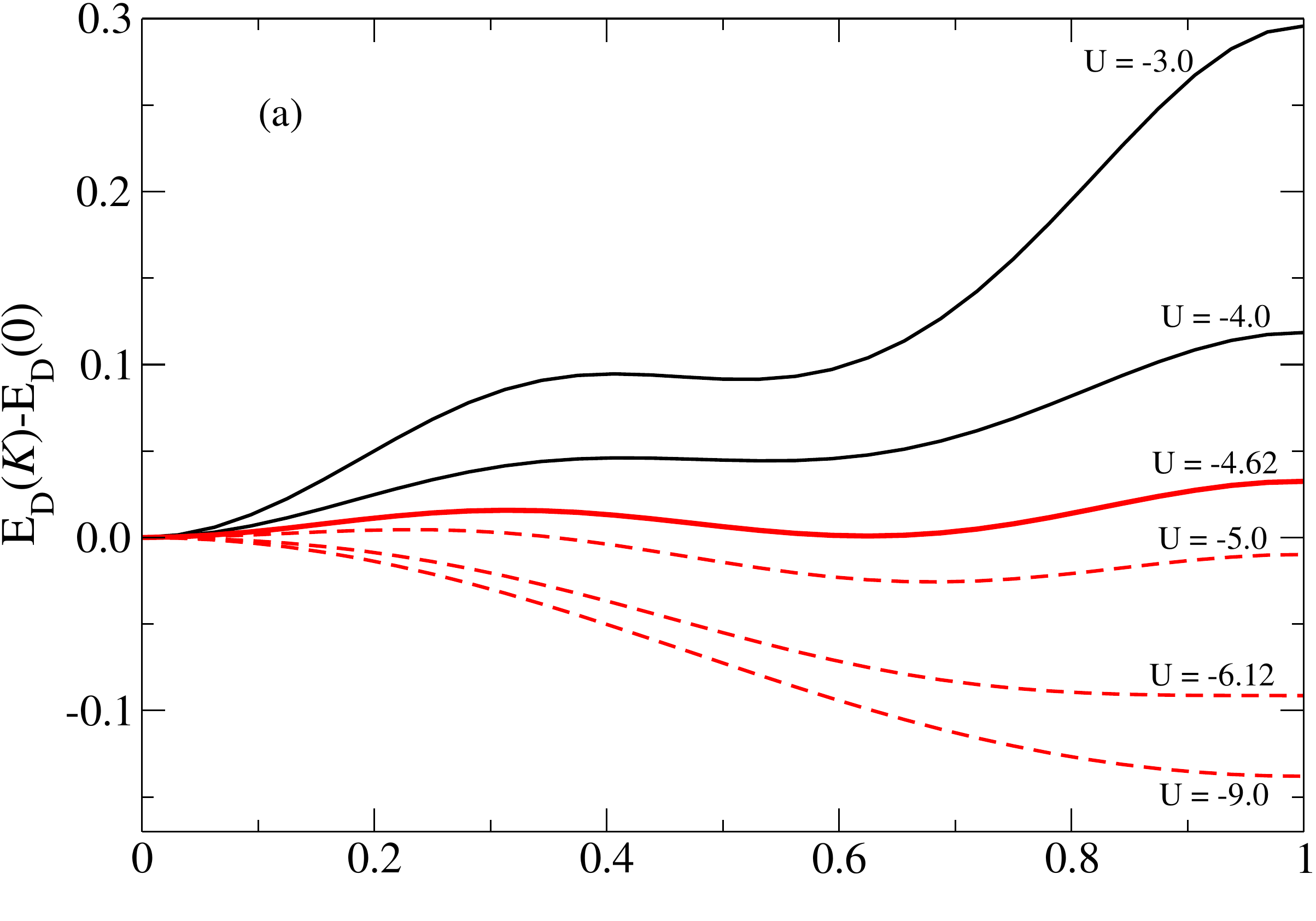} \centering
\includegraphics[width=0.9\columnwidth]{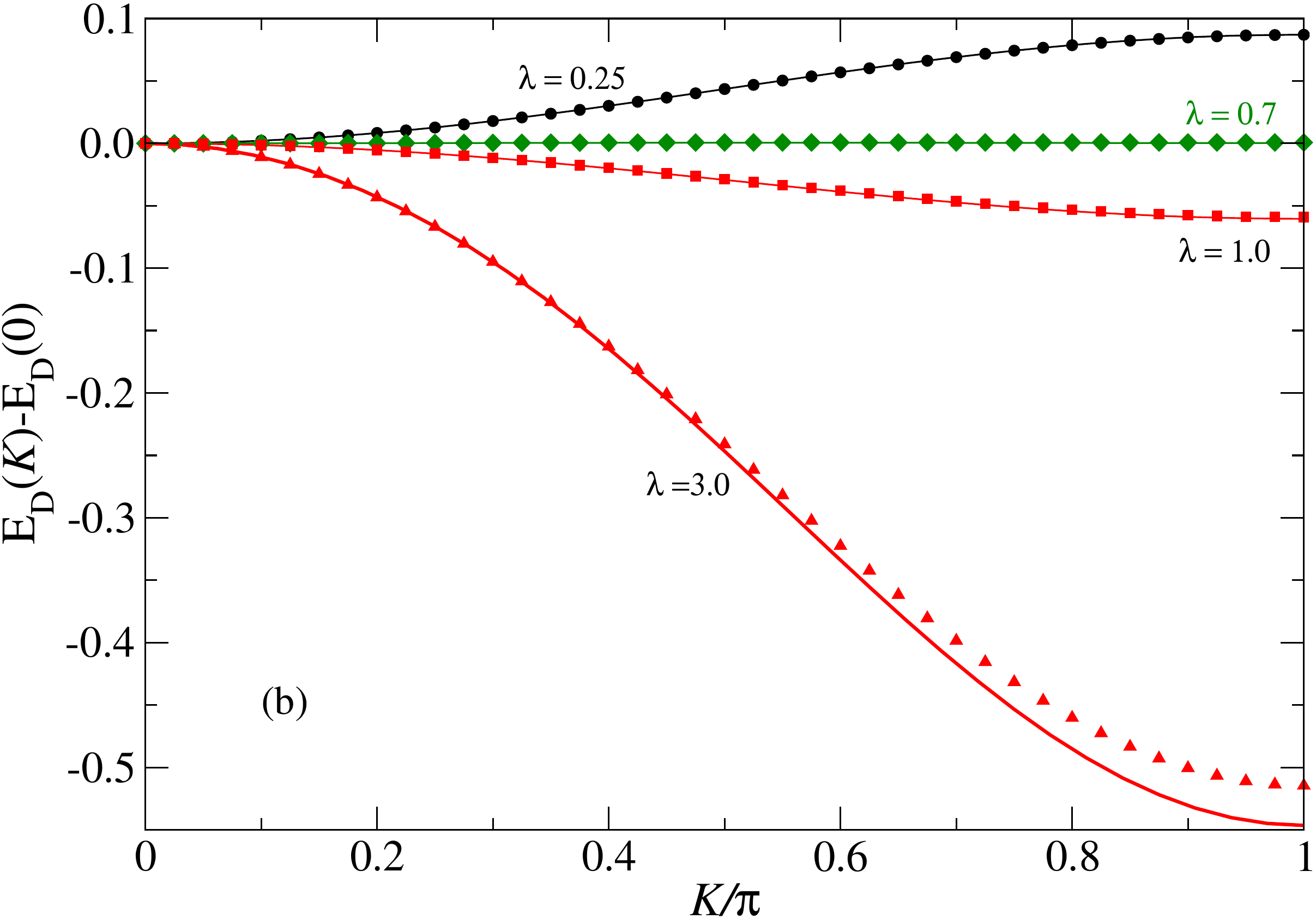}\caption{(color online)
Dimer dispersion $E_{D}(K)-E_{D}(0)$ for (a) $\lambda=1$ and various
values of $U$; and (b) $U=-30$ and various values of $\lambda$. In both cases
$t=1, \Omega=3$. The lines are the VED results and the symbols are the MA results. 
Note the sharp transitions of the GS momentum from $K_{\rm GS} = 0$ to $K_{\rm GS} >0$ 
in both cases.   \label{fig3}} \end{figure}

We now examine the properties of dimers formed when $U$ is
sufficiently large to balance the phonon-mediated repulsion and the
loss of kinetic energy. Figure \ref{fig3} shows the dimer
dispersion, $E_{D}(K)$ as a function of $U$ and $\lambda$, illustrating two unique
features of dimers arising from the SSH coupling. At low $|U|$
and/or $\lambda$, the dimer ground state has momentum $K_{GS} =
0$. As $\lambda$ and/or $|U|$ increases, there is a sharp transition
to a GS momentum $K > 0$. Fig. 3(a) shows that with increasing $|U|$,
the dimer dispersion develops a rather unusual shape with a second
local minimum appearing at a finite momentum. At $U\approx -4.62t$ this
minimum becomes degenerate with that at $K=0$, and the dimer
ground-state momentum jumps discontinuously to $K_{GS}\approx 0.6\pi$
and then continues to increase with increasing $|U|$. There is a
second sharp transition to $K_{GS}=\pi$ at $U=-6.12t$. These curves are
at a fixed $\lambda$ so the polaron dispersion is unchanged. The
change in the dimer dispersion (and in $K_{GS}$) is therefore due
to forcing the bound polarons closer, as $|U|$ increases.

In Fig. \ref{fig3}b we follow the evolution of $E_{D}(K)$ with
$\lambda$ for a fixed $U=-30t$. At small $\lambda$ we see a rather
heavy dimer with $K_{GS}=0$, as expected because in this limit the
non-interacting polarons are quite heavy and with  $E_P(k)$ increasing monotonically with $k$
 \cite{Dominic}. With increasing $\lambda$ the
effective dimer mass increases fast and the dispersion becomes flat. At a value $\lambda^* \approx 0.7$ the minimum jumps
discontinuously to $K_{GS}=\pi$. It stays there with further increase
in $\lambda$, but the bandwidth increases dramatically as the
phonon-mediated  pair-hopping terms become dominant, thus making the dimers
light at strong coupling.

Figure \ref{fig1} illustrates the locations of these sharp transitions
for the dimers in the extended Peierls-Hubbard model on the $U$-$\lambda$ phase diagram. To the best
of our knowledge, this is the first observation of such sharp
transitions of the two--polaron ground state. They never occur in Holstein or Fr\"{o}hlich 
models, where the bipolarons always have $K_{GS}=0$ \cite{BT,HK,M1}.

The second unique feature illustrated in Figure 3 is the flattening of
the dimer dispersion at $\lambda = \lambda^\ast \approx 0.7$,
suggestive of {\em self-trapping}: here the dimers are essentially
localized even though the single polarons have finite bandwidth. This
behavior can be understood qualitatively as follows. For small
$\lambda$, the polaron dispersion is dominated by its NN hopping. A
large $|U|$ can bind the polarons only when they are on adjacent
sites. When acting on such a configuration, NN hopping moves the
particles two sites apart to an energetically expensive configuration.
As a result, the effective dimer dispersion acquires a term $\sim
-t^2/|U|\cos(K)$, which favors $K_{GS}=0$ (see Supplementary Information). On the other
hand, at finite $\lambda$, the ``pair-hopping'' process moving the NN
pair as a whole also becomes active and contributes a term of order
$2t_2 \cos (K)$ to the dimer dispersion (see Supplementary Information); this term
favors $K_{GS}=\pi$. At $\lambda = \lambda^*$ the two terms cancel
 and the bandwidth collapses. However, numerical simulations
cannot guarantee that the bandwidth is precisely zero, and we do not
have theoretical arguments why the longer range hopping should also
vanish at $\lambda^*$. This is why we prefer to use the more conservative
term of (quasi) self-trapping.


Before concluding, we highlight another accomplishment of this work,
demonstrated by Figs. 1 and 3, namely the successful generalization of
the MA approximation to bipolaron{-type} problems. The variational space we
implemented here (see Supplementary Information) is designed, by construction, to describe
strongly bound polarons. Indeed, the MA predictions are in
quantitative agreement with VED in this limit. The MA results are also
qualitatively correct near $U_C(\lambda)$ (not shown), but its
accuracy is much poorer for weakly bound polarons. A suitable increase
of the variational space is necessary to improve the accuracy of the
MA approximation for weakly bound states. This can be done in a rather
straightforward way and promises to establish MA as an equally
valuable and efficient method for the study of bipolarons as it is for
polarons.

{\em Discussion:}
To summarize, we showed that dressing interacting particles by phonons
through SSH/Peierls couplings leads to very rich two-polaron physics,
qualitatively different from what is known for conventional polaron
models. We showed that for bare particles with the statistics of
identical fermions or of hard-core bosons, the phonon-mediated
interactions are {repulsive}, contradicting the conventional view
that phonons act as ``glue'' for quasiparticles. We showed that
the ``pair-hopping'' terms, which are also mediated by phonon-exchange
and can only arise in models with phonons modulating the particle hopping, play a major role, leading to
{\it sharp transitions} of the {bound dimer's} ground state. We also observe
the collapse of the dimer's dispersion at phonon coupling strength
$\lambda^*$ where the single polarons are mobile, suggestive of a
self-trapping transition.

As discussed in the previous section, all these new observations rest on the interplay of two generic features: hard-core statistics of bare particles and off-diagonal, hopping-dependent particle-phonon couplings.
As such, these results apply to a wide range of systems and have far-reaching implications for complex quantum
systems of interacting dressed particles. 
 \revision{The Hamiltonians considered here describe the interactions of small excitons coupled to phonons, particularly relevant to molecular crystals and organics semiconductors \cite{materials,solar}.
 Moreover, the hopping-dependent interactions with phonons, such as the one described by Eq. (2), are generally present in all materials. They may not always be dominant but, because they lead to qualitatively distinct behaviour of the resulting dressed particles, our work raises an important question of how the interplay of the coupling terms in Eq. (2) with conventional phonon-induced interactions changes the dynamics of polarons. 
 As we showed in previous work \cite{polar-molecules2}, a perturbative admixture of the hopping-dependent interactions may lead to non-perturbative changes of the single polaron dispersion. If a similar effect happens for dimers or bipolarons at experimentally relevant interaction parameters, many of the long-standing questions in polaron physics must be re-visited to account for the hopping-dependent interactions with phonons. 
   }

In addition, our results suggest that soft-core bosons and/or singlet
fermions may form highly mobile bipolarons with sharp transitions even
in the limit of vanishing $U$; we are currently investigating this.
Also, many of these features are expected to apply to systems with
more particles. The ``pair-hopping'' terms must be equally important
for few-polaron ensembles, suggesting that the ground state of
few-polaron states must also exhibit sharp transitions and, perhaps,
localization (self-trapping).
%
%
%
%
%

{\em Acknowledgements}: This work was supported by NSERC of Canada and the Stewart Blusson Quantum Matter Institute. J. S.
acknowledges discussions with Ian Affleck. M. C. appreciates access to the
computing facilities of the DST-FIST (phase-II) project installed in the
Department of Physics, IIT Kharagpur, India.

{\em Author Contributions}: J. S., R. V. K. and M. B. designed the research project, J. S. and M. C. performed numerical simulations, J. S. and M. B. performed analytical calculations, J. S., C. P. J. A. and M. B. contributed to MA concepts, M. C. contributed to VED concepts, J. S., R. V. K. and M. B. interpreted results, all authors contributed to writing of the manuscript. 

{\em Competing financial interests:} The authors declare no competing financial interests.

\end{document}


\title{Supplementary Material for \\
``Phonon-mediated repulsion, sharp transitions and (quasi)self-trapping in the extended Peierls-Hubbard model''
\vspace{-0.9em}}

\date{\today}

\author{J. Sous}
\author{M. Chakraborty} 
\author{C. P. J. Adolphs}
\author{R. V. Krems}
\author{M. Berciu}
\maketitle

\section{Momentum Average (MA) Approximation - technical details}
The Momentum Average (MA) \cite{MA, MA2, MA3} approximation is a
non-perturbative quasi-analytical technique designed to solve the
equation of motion for the relevant Green's function $G(k,\omega) =
\langle k| (\omega - \mathcal{H} + i \eta)^{-1} |k\rangle$ in the
Bogoliubov-Born-Green-Kirkwood-Yvon (BBKGY) hierarchy. The hierarchy
consists of an infinite set of coupled equations which are impossible
to solve exactly. By neglecting exponentially small
contributions  in the expansion, one simplifies the equations of motion to a
form that is readily solvable numerically. The guide to approximating
the hierarchy follows from the variational meaning of MA:  essentially
one solves the problem in a variational subspace.

The choice of the variational space depends on the details of the
Hamiltonian and state(s) of interest \cite{MA2}. For the Holstein
model, a one-site phonon cloud suffices to  provides  accurate results for single
polarons \cite{MA, MA2} and for S0 bipolarons \cite{Clemens}. S0
bipolarons are single-site strongly bound bipolarons. Taken together with the local nature of the Holstein coupling, this explains why a one-site phonon cloud is accurate to describe such states. 
For the Edwards and SSH models, the coupling to phonons is non-local
and therefore a bigger cloud is required to yield accurate results. A
three-site phonon cloud MA has been shown to be very accurate for such models
\cite{Edwards, Dominic}.

In this work, we generalize MA to study strongly bound two-particle 
states in the extended Peierls/Su-Schrieffer-Heeger (SSH)--Hubbard
model. We derive the MA equations for two hardcore particles in a
three-site phonon cloud and allow the particles to be arbitrarily far
from the cloud but at most two sites apart from each other, if a cloud is present. Terms corresponding to the
particles being further than two sites apart  are expected to contribute significantly only to higher-energy states, if a two-particle state is strongly bound, which is the case of primary interest to us in this work. For weakly bound dimers, the variational space must be extended to include configurations where the particles are further apart. The variational space can be 
increased systematically until convergence is achieved.

To highlight the method we derive a few representative equations of motion used in
the MA formalism developed here \cite{follow-up}. This is achieved using Dyson's identity $\hat{G}(\omega) =
\hat{G}_{0}(\omega) + \hat{G}(\omega) \hat{V} \hat{G}_{0}(\omega)$ where $\hat{G}(\omega)=(\omega - \mathcal{H} + i \eta)^{-1}$, $\hat{G}_0(\omega)=(\omega - \mathcal{H}_0 + i \eta)^{-1}$ with $\mathcal{H}_0 = \mathcal{H}_p +
\mathcal{H}_{ph}$, and $\hat{V}$ is the bare particle - phonon coupling
term.

Consider  the two-particle propagator $G(K,1, n,\omega) = \langle
K,1|\hat{G}(w)\ket{K,n}$ defined for two-particle states $\ket{K,n} =
\sum_i \frac{e^{iK(R_i+na/2)}}{\sqrt{N}} c_i^\dagger c_{i+n}^\dagger
\ket{0}$ with the two particles $n\geq 1$ sites apart, and $a$ is the lattice constant. Using Dyson's identity and inserting a
resolution of the identity, its exact equation of motion can be written as:
\begin{eqnarray}
\EqLabel{J1}
G(K,1,n,\omega) &=& G_{0}(K,1,n,\omega) \nonumber \\ 
&+& \sum_{\eta} G_0(K,\eta, n,\omega)\langle K,1|\hat{G}(\omega)\hat{V}\ket{K,\eta}. \nonumber
\end{eqnarray}
Note that $G_{0}(K,1,n,\omega)$ can be calculated exactly analytically in one dimension \cite{Ring}.
Consider now $\hat{V}\ket{K,\eta}$. It consists of states with one phonon plus the particles $\eta \pm 1$ sites apart. Thus, the right-hand side of the exact equation of motion contains an infinite number of terms.  Because within MA we restrict the particles to be within two sites of each other when phonons are present, this simplifies the equation of motion to:
\begin{eqnarray}
\EqLabel{J2}
G(K,1,n,\omega) &=& G_{0}(K,1,n,\omega) \nonumber \\
&-&ge^{-iKa}G_0(K,2,n,\omega)F_1(-2,1)+gG_0(K,2,n,\omega)F_1(-1,1) \nonumber \\
&-&gG_0(K,2,n,\omega)F_1(0,1)+ge^{iKa}G_0(K,2,n,\omega)F_1(1,1) \nonumber \\
&-&ge^{-3iKa/2}G_0(K,3,n,\omega)F_1(-3,2) \nonumber \\
&-&g[e^{-3iKa/2}G_0(K,1,n,\omega)-e^{-iKa/2}G_0(K,3,n,\omega)]F_1(-2,2)\nonumber \\
&-&2igsin(Ka/2)G_0(K,1,n,\omega)F_1(-1,2) \nonumber \\
&+&g[e^{3iKa/2}G_0(K,1,n,\omega)-e^{iKa/2}G_0(K,3,n,\omega)]F_1(0,2) \nonumber \\
&+&ge^{3iKa/2}G_0(K,3,n,\omega)F_1(1,2),
\end{eqnarray}
where $F_1(m,n)$ is shorthand for $F_1(K,m,n,\omega)$ defined as
\begin{equation}
\EqLabel{J3} 
F_l(K,m,n,\omega) \equiv \sum_i \frac{e^{iKR_i}}{\sqrt{N}} \langle K,1| \hat{G}(\omega) c_{i+m}^\dagger c_{i+m+n}^\dagger b_i^{\dagger l}\ket{0},
\end{equation}
{\em i.e.} a  generalized one-site cloud propagator. 
By introducing other appropriate generalized propagators:
\begin{eqnarray}
\EqLabel{J4} 
&F_{l_1,l_2}(K,m,n,\omega) \equiv \sum_i \frac{e^{iKR_i}}{\sqrt{N}} \langle K,1| \hat{G}(\omega) c_{i+m}^\dagger c_{i+m+n}^\dagger b_i^{\dagger l_1} b_{i+1}^{\dagger l_2}\ket{0},& \\
\EqLabel{J5} 
&F_{l_1,l_2, l_3}(K,m,n,\omega) \equiv \sum_i \frac{e^{iKR_i}}{\sqrt{N}} \langle K,1| \hat{G}(\omega) c_{i+m}^\dagger c_{i+m+n}^\dagger b_{i-1}^{\dagger l_1} b_i^{\dagger l_2} b_{i+1}^{\dagger l_3}\ket{0},&
\end{eqnarray}
for two-site cloud and three-site cloud configurations respectively,
and repeatedly applying the Dyson's identity, one derives the MA
equations of motion for the the propagators in Eqs.
\eqref{J2}-\eqref{J5}. This linear system of coupled equations is
solved numerically and the propagator of interest $G(K,1, n,\omega)$
is computed. Dimer bound state properties such as $E_{D}(K)$ can be extracted
from the propagator.

By construction the MA approach developed here is designed to describe
strongly bound states accurately. Indeed, the comparison with Variational
Exact Diagonalization (VED) results shows that the MA approximation is accurate in this regime. By systematically extending the
variational subspace we expect to be able to study weakly-bound dimers as well.

\section{Perturbation theory (PT) results in the limit $\Omega \gg |t|, |g|$}

\subsection{Single-particle sector}

Let $\hat{P}$ be the projector onto the zero-phonon Hilbert subspace, which is spanned by the states $c_i^\dagger|0\rangle$, $\forall i$. The effective Hamiltonian in this subspace is, to second order in perturbation theory:
\begin{equation*}
\EqLabel{s1}
\hat{h}_1 = \hat{T} + \hat{P} \hat{V} {1\over {E_0 - {\cal H}_0}} \hat{V}\hat{P},
\end{equation*}
where $\hat{T} =-t \sum_{i}^{} (c_i^\dagger c_{i+1} + h.c.)$ is the bare kinetic energy, $\hat{V}$ is the bare particle - phonon coupling from Eq. (1) in the main text, and ${\cal H}_0 = {\cal H}_{\rm ph}$. The projection is straightforward to carry out and leads to:
\begin{equation}
\EqLabel{s2}
\hat{h}_1 = \hat{T} +  \hat{T}_2- {4g^2\over \Omega} \sum_{i}^{} \hat{n}_i,
\end{equation}
where $\hat{T}_2 = + t_2 \sum_{i}^{} (c_i^\dagger c_{i+2} + h.c.)$ is a phonon-mediated next-nearest-neighbor (NNN) hopping with $t_2 = g^2/\Omega= \lambda t/2$ (note the unusual sign). For convenience we define $\epsilon_0 ={ 4g^2\over \Omega}$.

Setting $a=1$, the polaron dispersion is, therefore,  $$E_P(k) = - \epsilon_0 - 2t \cos(k) + 2 t_2 \cos(2k).$$ It is straightforward to verify that if $t > 4t_2$, {\em i.e.} if $\lambda < {1\over 2}$, the polaron ground state (GS) momentum is 0. For $\lambda > {1\over 2}$, the polaron GS momentum is $k_P= \arccos {t\over 4 t_2}$, going asymptotically to ${\pi \over 2}$ as $\lambda\rightarrow \infty$. 

\subsection{Two-particle sector}

\begin{figure}[t]
    \centering
        \includegraphics[width=1.0\textwidth]{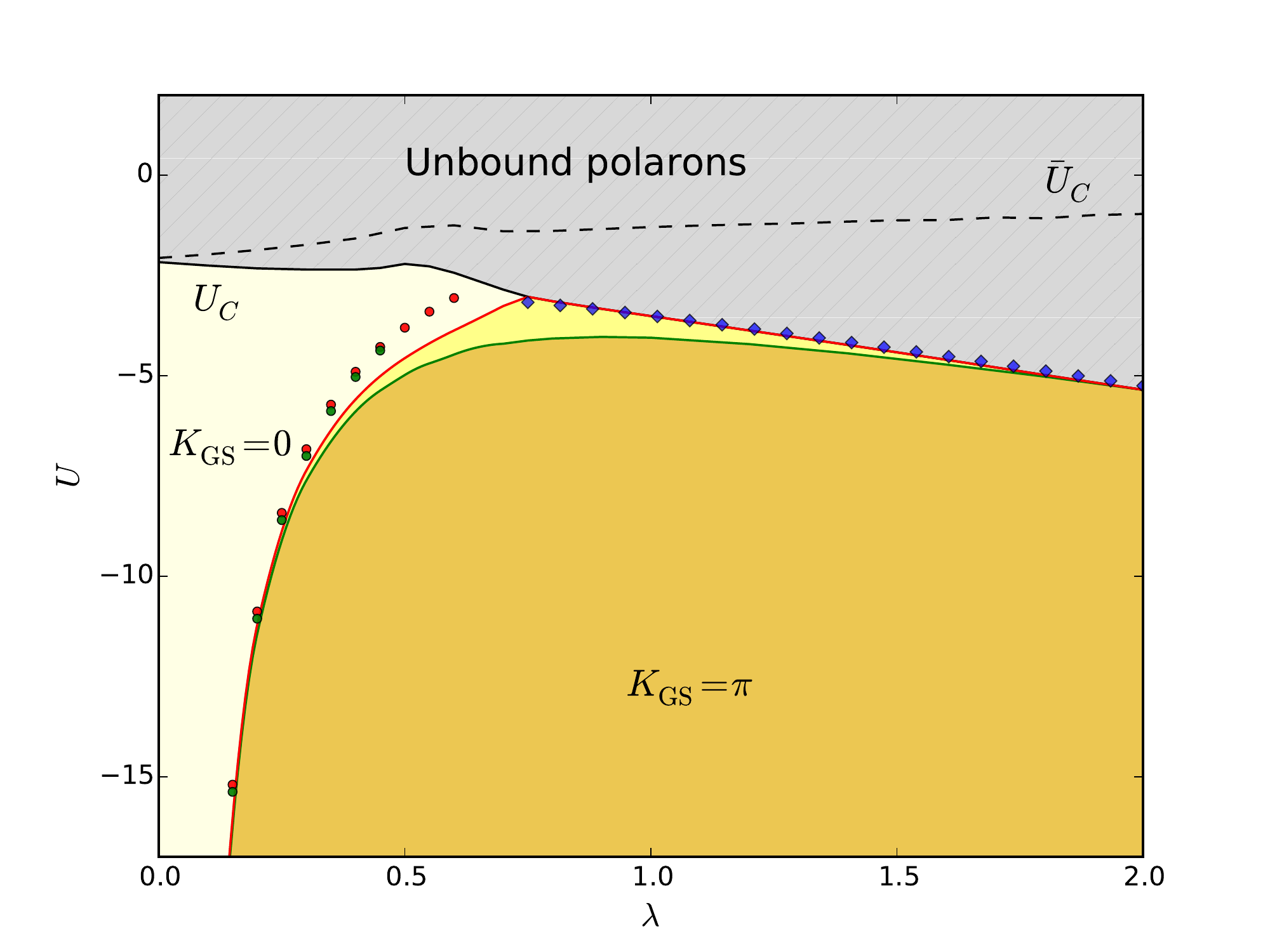}
   \caption{(color online) Two-polaron phase diagram for $t=1$, $\Omega=100$. The solid black line shows $U_C(\lambda)$ below which stable bound states form, while the dashed line shows $\bar{U}_C(\lambda)$ defined in the main text; the difference between the two is the strength of the phonon-mediated repulsion. The red and green lines mark the sharp transitions of the bound dimer's GS. The lines are the VED results and the symbols of the corresponding color are the MA results. The diamond symbols are the PT results of Eq. \eqref{s4}. The area between the red and green curves corresponds to a dimer bound state with $0 < K_{\rm GS} < \pi$.
    \label{fig:s1}}
\end{figure}

Repeating the projection onto the two-bare particle -- zero-phonon subspace spanned by the states $c^\dagger_i c^\dagger_{i+n}|0\rangle$, $\forall n \ge 1, i$, we find:
\begin{equation}
\EqLabel{s3}
\hat{h}_2 = \hat{T} + \hat{T}_2- \epsilon_0\sum_{i}^{} \hat{n}_i + \tilde{U} \sum_{i}^{} \hat{n}_i \hat{n}_{i+1},
\end{equation}
where $\tilde{U} = U + \epsilon_0$; $U$ is the bare nearest-neighbor (NN) interaction.

The two-polaron bound state dispersion can be calculated numerically, either using direct diagonalization for a large-enough chain, or using the Equation-of-Motion (EOM) approach. We briefly present the latter and then use it to obtain analytical solutions in some specific limits.

We define $|K,n\rangle = \sum_{i}^{} {e^{iK(R_i+{n\over 2})}\over \sqrt{N}} c^\dagger_i c^\dagger_{i+n}|0\rangle$, $\forall n \ge 1$, and the propagators $g(n) \equiv \langle K, 1| \hat{G}(\omega) |K,n\rangle$, where $\hat{G}(\omega) = ( \omega + i \eta - \hat{h}_2)^{-1}$ is the resolvent of interest. The bound state energy (once a bound state appears) is at the lowest discrete pole of these propagators.  Using the identity $\hat{G}(\omega) ( \omega + i \eta - \hat{h}_2)^{-1}=1$, we generate the EOM:
 \begin{align*}
&(\omega + i \eta -\tilde{U} +2\epsilon_0-\beta_K) g(1)  = 1 - \alpha_K g(2) + \beta_K g(3)\\
   &(\omega + i \eta +2\epsilon_0) g(2)   = -\alpha_K [g(1)+g(3)]+ \beta_K g(4)
 \end{align*}
 and for any $n\ge 3$,
 \begin{align*}
   (\omega + i \eta +2\epsilon_0) g(n)   = -\alpha_K [g(n-1)+g(n+1)]\\ + \beta_K [g(n-2)+g(n+2)].
 \end{align*}
   Here,  $\alpha_K = 2t \cos ({K\over 2})$, $\beta_K = 2t_2 \cos(K)$.

 The physically acceptable analytical solution for recurrence
 relations of this type is available in \cite{math}, however it is rather
 complicated and its poles cannot be extracted analytically. A general
 solution can be found numerically.

 There are two cases that can be solved rather easily analytically, namely (i) if $\beta_K
 = 0$, and (ii) if $\alpha_K=0$. The first is realized when $t_2=0$,
 and can be used as an indication of physics at very weak
 couplings $\lambda \rightarrow 0$. In this case the recurrence
 relation becomes trivial. For any $n\ge 2$ we have $g(n) =
 z(K,\omega) g(n-1)$ where
 $$ z(K,\omega) = {1\over 2 \alpha_K} \left[- \tilde{w}
   +\sqrt{\tilde{w}+2\alpha_K}\sqrt{\tilde{w}-2\alpha_K}\right]
 $$ and $\tilde{\omega} \equiv (\omega + i \eta +2\epsilon_0)$. This results in:
 $$
g(K,n,\omega) = 2 {[z(K,\omega)]^{n-1} \over \tilde{w} - 2 \tilde{U} + \sqrt{\tilde{w}+2\alpha_K}\sqrt{\tilde{w}-2\alpha_K}}
 $$
for which exists a line cut that indicates a continuum (the two-particle continuum) for $|\tilde{\omega}|\le 2\alpha_K$. A bound dimer appears if and only if there is a discrete pole at $\omega=E_{D}(K)$ such that a) $E_{D}(K) + 2\epsilon_0< - 2\alpha_K$ ( the pole is below the continuum) and b) $E_{D}(K) + 2\epsilon_0 - 2 \tilde{U} + \sqrt{E_{D}(K)+2\epsilon_0+2\alpha_K}\sqrt{E_{D}(K)+2\epsilon_0-2\alpha_K} =0 $ (the pole condition). It follows that the solution is
$$
E_{D}(K) = - 2\epsilon_0 + \tilde{U} + {\alpha_K^2\over \tilde{U}},
$$
which is below the continuum if and only if $\tilde{U} \le - \alpha_K$. In particular, this requires $\tilde{U} < -2t$ for a bound state to emerge in the entire Brillouin zone, even at $K=0$. This finite bare NN attraction is necessary in order to compensate for the lost kinetic energy, when the two particles bind.

The second accessible analytical solution is for $\alpha_K=0$, which is valid for $K=\pi$ irrespective of the value of $t$, and therefore will give the exact PT bound dimer solution at the edge of the Brillouin zone. The solution follows as before, and it results in:
$$
g(K,1,\omega) = {2\over \tilde{\omega} -2\tilde{U} + 4t_2 + \sqrt{\tilde{\omega} - 4t_2}\sqrt{\tilde{\omega} + 4t_2}}
$$
for which, repeating the analysis, we find that a discrete bound state appears if and only if $\tilde{U} <0\rightarrow U < -2\lambda t$, and its energy is $E_{D}(\pi) = -2\epsilon_0 + \tilde{U}- 2t_2 + {4t_2^2\over \tilde{U}-2t_2}$.

Note that the definition we employ for a {\em stable} bound state is not one that lies below the continuum at its momentum, but one that lies below the edge (lowest overall possible value) of the continuum. If this more stringent condition is not met, coupling to other fields (acoustic phonons, photons, ...) would allow transitions from this discrete state into the bottom of the continuum, so the dimer state is not truly stable.

To calculate $U_C$ according to this definition, we compute the lower edge of the continuum. Given the single polaron dispersion $E_P(k)$ (calculated in the previous section), and noting that at a given total momentum $K$ the two-polaron continuum start at $\min_q [E_P(K-q) +E_P(q)]$, one can easily solve for the lower edge of the continuum, $E_c$. One finds the result $E_c= 2 \min_k E_P(k)$, {\em i.e.} twice the polaron GS energy.

  Requiring that $E_{D}(\pi) < E_c$ leads to
\begin{equation}
\EqLabel{s4}
U_C(\lambda) = - 2 \lambda t - t- {t\over 2\lambda} + {\cal O}\left({1\over \lambda^2}\right).
\end{equation}
This should agree with numerical results for sufficiently large $\lambda$ where the GS momentum of the bound state approaches $K_{GS}=\pi$. However, note that $\lambda$ should still be small enough such that PT remains valid. For example, for $\Omega =100, t=1$ this would require $\lambda < 2 \rightarrow g < 10$ such that $g/\Omega \ll 1$. For stronger couplings one needs to go to higher order(s) in perturbation theory.

Indeed, Fig. \ref{fig:s1} shows good agreement at moderate and large  $\lambda \le 2$ between $U_C(\lambda)$ and this analytical prediction of Eq. \eqref{s4}.